\documentclass[11pt,twoside,onecolumn]{article}
\usepackage[]{latexsym,amsmath,amssymb}
\usepackage[]{epsfig}
\newcommand{\be}{\begin{eqnarray}}
\newcommand{\ee}{\end{eqnarray}}

\title{\bf Generalized Uncertainty Principle, Extra-dimensions and
Holography}
\author{Fabio Scardigli$^a$\thanks{Corresponding author. Address for the
correspondence: Via Europa 20, 20097 S. Donato Milanese, Milano, Italy.
E-mail: fabio@itp.unibe.ch}
$\ $and
Roberto Casadio$^b$\thanks{E-mail: casadio@bo.infn.it}
\\
\\
{\em $^a$Institute for Theoretical Physics, University of Bern,}
\\
{\em Sidlerstrasse~5, 3012 Bern, Switzerland}
\\
\null
\\
{\em $^b$Dipartimento di Fisica, Universit\`a di
Bologna and I.N.F.N., Sezione di Bologna,}\\
{\em via Irnerio~46, 40126 Bologna, Italy}}
\begin{document}
\maketitle
\begin{abstract}
We consider Uncertainty Principles which take into account the role
of gravity and the possible existence of extra spatial dimensions.
Explicit expressions for such Generalized Uncertainty Principles
in $4+n$ dimensions are given and their holographic properties
investigated.
In particular, we show that the predicted number of degrees of
freedom enclosed in a given spatial volume matches the holographic
counting only for one of the available generalizations and without
extra dimensions.
\par
\null
\par
\textit{PACS 04.60 - Quantum theory of gravitation.}
\end{abstract}
\raggedbottom
\setcounter{page}{1}
\section{Introduction}
\setcounter{equation}{0}
During the last years many efforts have been devoted to clarifying
the role played by the existence of extra spatial dimensions in the
theory of gravity \cite{add,RS}.
One of the most interesting predictions drawn from the theory is that
there should be measurable deviations from the $1/r^2$ law of Newtonian
gravity at short (and perhaps also at large) distances.
Such new laws of gravity would imply modifications of those Generalized
Uncertainty Principles (GUP's) designed to account for gravitational
effects in the measure of positions and energies.
\par
On the other hand, the holographic principle is claimed to apply
to all of the gravitational systems.
The existence of GUP's satisfying the holography in four dimensions
(one of the main examples is due to Ng and Van~Dam \cite{ngvd}) led
us to explore the holographic properties of the GUP's extended to the
brane-world scenarios.
The results, at least for the examples we considered, are quite
surprising.
The expected holographic scaling indeed seems to hold only in four
dimensions, and only for the Ng and van~Dam's GUP.
When extra spatial dimensions are admitted, the holography is destroyed.
This fact allows two different interpretations:
either the holographic principle is not universal and does not apply
when extra dimensions are present; or, on the contrary, we take seriously
the holographic claim in any number of dimensions, and our results are
therefore evidence against the existence of extra dimensions.
\par
In Section~\ref{lgup} we analyze GUP's obtained by linearly combining
quantum mechanical expressions with general relativistic bounds \cite{FS};
in Section~\ref{ng} we repeat the same analysis for the type of GUP's
discussed in Refs.~\cite{ngvd} and in Section~\ref{conc} we comment on
our results.
The four-dimensional Newton constant is denoted by $G_{\rm N}$ throughout
the paper.
\section{Linear GUP's from micro black holes}
\setcounter{equation}{0}
\label{lgup}
In this Section we derive GUP's via a micro black hole gedanken
experiment, following closely the content of Ref.~\cite{FS}
which we then generalize to space-times with extra dimensions.
\subsection{GUP in four dimensions}
\label{gup_4}
When we measure a position with precision of order $\Delta x$,
we expect quantum fluctuations of the metric field around the
measured position with energy amplitude
\be
\Delta E \sim \frac{\hbar \, c}{2\, \Delta x}
\ .
\label{dE}
\ee
The Schwarzschild radius associated with the energy $\Delta E$,
\be
R_{\rm S}=\frac{2\,G_{\rm N}\,\Delta E}{c^4}
\ ,
\label{Rs4}
\ee
falls well inside the interval $\Delta x$ for practical cases.
However, if we wanted to improve the precision indefinitely,
the fluctuation $\Delta E$ would grow up and the corresponding
$R_{\rm S}$ would become larger and larger, until it reaches the same
size as $\Delta x$.
As it is well known, the critical length is the Planck length,
\be
\Delta x = R_{\rm S} \quad \Rightarrow \quad
\Delta x  = \left(\frac{G_{\rm N}\,\hbar}{c^3}\right)^{1/2}
\equiv \ell_{\rm p}
\ ,
\label{lp}
\ee
and the associated energy is the Planck energy
\be
\epsilon_{\rm p}\equiv {\hbar\,c\over 2\,\ell_{\rm p}}
={1\over 2}\left({\hbar\,c^5\over G_{\rm N}}\right)^{1/2}
\ .
\label{ep}
\ee
If we tried to further decrease $\Delta x$, we should concentrate
in that region an energy greater than the Planck energy, and this
would enlarge further the Schwarzschild radius $R_{\rm S}$, hiding more
and more details of the region beyond the event horizon of the
micro hole.
The situation can be summarized by the inequalities
\be
\Delta x \gtrsim \left\{ \begin{array}{lcl}
\frac{\hbar\,c}{2\,\Delta E}
&{\rm for}&
\Delta E < \epsilon_{\rm p}
\\
\\
\frac{2\,G_{\rm N}\,\Delta E}{c^4}
&{\rm for}&
\Delta E > \epsilon_{\rm p}
\ .
\end{array}
\right.
\ee
which, if combined linearly, yield
\be
\Delta x \gtrsim \frac{\hbar\,c}{2\,\Delta E}
+ \frac{2\,G_{\rm N}\,\Delta E}{c^4}
\ .
\label{gup1}
\ee
This is a generalization of the uncertainty principle to cases in which
gravity is important, i.e.~to energies of the order of $\epsilon_{\rm p}$.
We note that the minimum value of $\Delta x$ is reached for
$(\Delta E)_{\rm min}=\epsilon_{\rm p}$ and is given by
$(\Delta x)_{\rm min}=2\,\ell_{\rm p}$.
\subsection{GUP with $n$ extra dimensions}
\label{gup_n}
We shall now generalize the procedure outlined in the previous
Subsection to a space-time with $4+n$ dimensions, where $n$ is
the number of space-like extra dimensions.
The first problem we should address is how to relate the gravitational
constant $G_{\rm N}$ in four dimensions with the one in $4+n$, henceforth
denoted by $G_{(4+n)}$.
\par
This of course depends on the model of space-time with extra
dimensions we consider.
Models appeared in the literature in recent years belong mostly
to two scenarios:
\begin{itemize}
\item
the Arkani-Hamed--Dimopoulos--Dvali (ADD) model \cite{add},
where the extra dimensions are compact and of size $L$;
\item
the Randall--Sundrum (RS) model \cite{RS}, where the extra
dimensions have an infinite extension but are warped by a
non-vanishing cosmological constant.
\end{itemize}
A feature shared by (the original formulations of) both
scenarios is that only gravity propagates along the $n$ extra
dimensions, while Standard Model fields are confined on
a four-dimensional sub-manifold usually referred to as the
{\em brane-world\/}.
\par
In the ADD case the link between $G_{\rm N}$ and $G_{(4+n)}$ can be fixed
by comparing the gravitational action in four dimensions with
the one in $4+n$ dimensions.
The space-time topology in such models is
$\mathcal{M}=\mathcal{M}^{4}\otimes \Re ^{n}$, where
$\mathcal{M}^{4}$ is the usual four-dimensional space-time and
$\Re^n$ represents the extra dimensions of finite size $L$.
The space-time brane has no tension and therefore the action
$S_{(4+n)}$ can be written as
\be
S_{(4+n)}=\frac{c^4}{16\,\pi\,G_{(4+n)}}\,
\int_{\mathcal{M}^{4}\otimes \Re^{n}}d^{4+n}x\,\sqrt{-g}\,R
\sim
\frac{c^4}{16\,\pi\,G_{(4+n)}}\,\int_{\mathcal{M}^{4}}
d^{4}x\,\sqrt{-\tilde{g}}\,L^n\,\tilde{R}
\ ,
\ee
where $\tilde{R}$, $\tilde{g}$ are the projections on
$\mathcal{M}^{4}$ of $R$ and $g$.
Here $L^n$ is the ``volume'' of the extra dimensions and we omitted
unimportant numerical factors.
On comparing the above expression with the purely four-dimensional
action
\be
S_{(4)}=\frac{c^4}{16\,\pi\,G_{\rm N}}
\int_{\mathcal{M}^{4}}d^{4}x\,\sqrt{-\tilde{g}}\,\tilde{R}
\ ,
\ee
we obtain
\be
G_{(4+n)}\sim G_{\rm N}\,L^n
\ .
\label{G4n}
\ee
\par
The RS models are more complicated.
It can be shown \cite{RS} that for $n=1$ extra dimension we have
$G_{(4+n)}=\sigma ^{-1}\,G_{\rm N}$, where $\sigma$ is the brane tension
with dimensions of $length^{-1}$ in suitable units.
The gravitational force between two point-like masses $m$ and $M$
on the brane is now given by
\be
F=G_{\rm N}\,\frac{m\,M}{r^2}\,
\left(1+\frac{e^{-\sigma r}}{\sigma^2 r^2}\right)
\ ,
\label{FRS}
\ee
where the correction to Newton law comes from summing over the extra
dimensional graviton modes in the graviton propagator \cite{RS}.
However, since Eq.~(\ref{FRS}) is obtained by perturbative calculations,
not immediately applicable to a non-perturbative structure such as a
black hole, we shall consider only the ADD scenario in this paper.
To be more precise, from table-top tests of the gravitational force
one finds that $n\ge 2$ in ADD \cite{add,Gund}.
On the other hand, black holes with mass $M\ll\sigma^{-1}$ are likely
to behave as pure five-dimensional in RS \cite{katz}, therefore
our results for $n=1$ should apply to such a case.
\par
In order to proceed as in the previous Section, we now need a
formula for the Schwarzschild radius in $4+n$ dimensions. This can
be obtained, heuristically, from the gravitational force law in
$4+n$ dimensions \cite{Arg} as determined by Gauss theorem, \be
F=G_{(4+n)}\,\frac{m\,M}{r^{2+n}} \ . \ee Therefore, the total
energy of a particle of mass $m$ in the gravitational field of the
source $M$ is given by \be E_{\rm
tot}=\frac{1}{2}\,m\,v^2-G_{(4+n)}\,\frac{m\,M}{r^{1+n}} \ , \ee
and the escape velocity is \be
v^2_f=2\,G_{(4+n)}\,\frac{M}{r^{1+n}} \ . \ee Requiring $v_f=c$,
we obtain for the Schwarzschild radius \be
R_{(4+n)}=\left(\frac{2\,G_{(4+n)}\,M}{c^2}\right)^{\frac{1}{1+n}}
\ . \label{Rs4n} \ee We shall show in the next Section that an
exact calculations based on the higher dimensional Schwarzschild
solution \cite{MP} just modifies this result by numerical factors.
\par
In the following, we shall just consider micro black holes with
$R_{(4+n)}\ll L$, so as to avoid the complications that are expected
when the Schwarzschild radius approaches the compactification length
\cite{ch}.
Moreover, the opposite case ($R_{(4+n)}\gg L$) would imply
the complete non-observability of extra dimensions, hidden beyond the
event horizon.
\par
The radius $R_{(4+n)}$ is related to $R_{(4)}\equiv R_{\rm S}$ as given
in Eq.~(\ref{Rs4}) according to
\be
R_{(4+n)}=\left(\frac{2\,G_{(4+n)}\,M}{c^2}\right)^{\frac{1}{1+n}}
=\left(\frac{2\,G_{\rm N}\,M\,L^n}{c^2}\right)^{\frac{1}{1+n}}=
R_{\rm S}^{\frac{1}{1+n}}\,L^{\frac{n}{1+n}}
\ ,
\ee
and, from $R_{(4+n)}\ll L$, we can infer the following inequalities:
\be
&&R_{(4+n)} \ll L \quad \Rightarrow \quad
R_{(4+n)}^{1+n} \ll L^{1+n} \quad \Rightarrow
\quad R_{\rm S}\,L^n\,\ll L^{1+n} \quad \Rightarrow
\quad R_{\rm S} \ll L
\ ;
\\
&&R_{\rm S} \ll L \quad \Rightarrow \quad R_{\rm
S}^{\frac{n}{n+1}} \ll L^{\frac{n}{n+1}} \quad \Rightarrow \quad
R_{\rm S} \ll L^{\frac{n}{n+1}}\,R_{\rm S}^{\frac{1}{n+1}} \quad
\Rightarrow \quad R_{\rm S} \ll R_{(4+n)} \ . \ee
Therefore,
\be
R_{\rm S} \ll R_{(4+n)} \ll L
\ee
and the Schwarzschild radius of
a black hole in a space-time with extra dimensions is greater than
in four dimensions \cite{Arg,ch}.
\par
Since measurements can be performed only on the brane, to the
uncertainty $\Delta x$ in position we can still associate an
energy given by Eq~(\ref{dE}).
The corresponding Schwarzschild radius is now given by Eq.~(\ref{Rs4n})
with $M=\Delta E/c^2$ and the critical length such that
$\Delta x=R_{(4+n)}$ is the Planck length in $4+n$ dimensions,
\be
\Delta x=
\left(\frac{G_{(4+n)}\,\hbar}{c^3}\right)^{\frac{1}{2+n}}
=\left(\frac{G_{\rm N}\,\hbar}{c^3}\,L^n\right)^{\frac{1}{2+n}}
=\left(\ell_{\rm p}^2\,L^n\right)^{\frac{1}{n+2}}
\equiv \ell_{(4+n)}
\ .
\ee
The energy associated with $\ell_{(4+n)}$ is analogously
the Planck energy in $4+n$ dimensions,
\be
\epsilon_{(4+n)}=
\frac{1}{2}\left(\frac{\hbar\,c^5}{G_{\rm N}}\frac{\hbar^n\,c^n}{L^n}
\right)^{\frac{1}{n+2}}=
\frac{1}{2}\,\left[4\,\epsilon_{\rm p}^2\,
\left(\frac{\hbar\,c}{L}\right)^n\right]^{\frac{1}{n+2}}
\ ,
\ee
where $\epsilon_{\rm p}$ is the Planck energy in 4 dimensions
given in Eq.~(\ref{ep}).
\par
It is reasonable to assume that $\ell_{\rm p}\ll L$, otherwise the
extra dimensions would not have a classical space-time
structure.
We then have
\be
\ell_{\rm p} \ll L \quad \Rightarrow \quad \ell_{\rm p}^n \ll L^n
\quad \Rightarrow \quad
\ell_{\rm p}^{2+n} \ll \ell_{\rm p}^2\,L^n =\ell_{(4+n)}^{2+n}
\quad \Rightarrow \quad \ell_{\rm p} \ll \ell_{(4+n)}
\ .
\ee
Further, we can also prove that
\be
\ell_{\rm p} \ll L \quad \Rightarrow \quad
\ell_{\rm p}^2 \ll L^2 \quad \Rightarrow \quad
\ell_{(4+n)}^{2+n}=\ell_{\rm p}^2\,L^n \ll L^{2+n}
\quad \Rightarrow \quad
\ell_{(4+n)} \ll L
\ .
\ee
Summarizing, from $\ell_{\rm p}\ll L$ we obtain
\be
\ell_{\rm p} \ll \ell_{(4+n)} \ll L
\ ,
\ee
and correspondingly
\be
\frac{\hbar\,c}{2\,L} \ll \epsilon_{(4+n)} \ll \epsilon_{\rm p}
\ ,
\ee
so that the Planck energy threshold, where quantum gravity
phenomena become important, is lowered by the existence
of extra dimensions \cite{add}.
Finally, we can check the inequalities among $\ell_{\rm p}$,
$R_{\rm S}$, $R_{(4+n)}$, and $\ell_{(4+n)}$.
We can easily prove that
\be
\ell_{(4+n)}<R_{(4+n)}
\ ,
\ee
since
\be
\ell_{(4+n)}<R_{(4+n)} \quad \Leftrightarrow \quad
\left(\ell_{\rm p}^2\,L^n\right)^{\frac{1}{n+2}}
< \left(R_{\rm S}\,L^n\right)^{\frac{1}{n+1}}
\quad \Leftrightarrow
\quad \ell_{\rm p}^{n+2}\,\ell_{\rm p}^n < R_{\rm S}^{n+2}\,L^n
\ ,
\ee
and the last inequality holds by virtue of $\ell_{\rm p}<R_{\rm S}$ and
$\ell_{\rm p}<L$.
We are therefore left with two possible chains of inequalities,
\be
&&\quad \quad \ell_{\rm p} \quad < \quad R_{\rm S} \quad < \quad
\ell_{(4+n)} \quad < \quad R_{(4+n)} \quad < \quad L
\\
&&\quad \quad \ell_{\rm p} \quad < \quad \ell_{(4+n)} \quad <
\quad R_{\rm S}  \quad < \quad R_{(4+n)} \quad < \quad L
\ ,
\ee
and, in general, it is not possible to tell if $R_{\rm S}< \ell_{(4+n)}$
or $R_{\rm S}> \ell_{(4+n)}$.
\par
Now, let us come back to the GUP.
The argument goes precisely as in four dimensions and one therefore
obtains the following inequalities
\be
\Delta x \gtrsim \left\{ \begin{array}{lcl}
\frac{\hbar\,c}{2\,\Delta E}
&{\rm for}&
\Delta E < \epsilon_{(4+n)}
\\
\\
\left(\frac{2\,G_{(4+n)}\,\Delta E}{c^4}\right)^{\frac{1}{n+1}}
&{\rm for}&
\Delta E > \epsilon_{(4+n)}
\ .
\end{array}
\right.
\ee
Combining linearly the previous inequalities, we obtain
\be
\Delta x \gtrsim \frac{\hbar\,c}{2\,\Delta E} +
\left(\frac{2\,G_{(4+n)}\Delta E}{c^4}\right)^{\frac{1}{n+1}}
\ ,
\label{gup2}
\ee
which is a straightforward generalization of Eq.~(\ref{gup1}) to
the case with $n$ extra dimensions.
The minimum value for $\Delta x$ is now reached when
$(\Delta E)_{\rm min}=(1+n)^{\frac{1+n}{2+n}}\epsilon_{\rm p}$
and we then have
\be
(\Delta x)_{\rm min}=
\left[(1+n)^{-\frac{1+n}{2+n}}+(1+n)^{\frac{1}{2+n}}\right]\,
\ell_{(4+n)}
\ .
\label{dx4n}
\ee
\subsection{Holographic properties}
\label{lholo}
In this Subsection, we investigate the holographic properties of the
GUP's which we have proposed this far.
We shall estimate the number of degrees of freedom  $n(V)$ contained
in a spatial volume (cube or ``hypercube'') of size $l$.
The holographic principle claims that $n(V)$ scales as the area of the
(hyper-)surface enclosing the given volume, that is
$(l/\ell_{\rm p})^{2+n}$ in $4+n$ dimensions.
\par
For the GUP's considered in the previous Subsections we find:
\begin{description}
\item[a)]
for the four-dimensional GUP in Eq.~(\ref{gup1}), this scaling does
not occur.
In fact, $(\Delta x)_{\rm min}\sim\ell_{\rm p}$ and a cube of side $l$
contains a number of degrees of freedom equal to
\be
n(V)\sim \left(\frac{l}{\ell_{\rm p}}\right)^3
\ .
\ee
\item[b)]
for the GUP in $4+n$ dimensions of Eq.~(\ref{gup2}), the minimum value
for $\Delta x$ is given in Eq.~(\ref{dx4n}) and, a part from numerical
factors, we see that the holographic scaling again does not hold,
\be
n(V)\sim \left(\frac{l}{(\Delta x)_{\rm min}}\right)^{3+n}
\sim \left(\frac{l}{\ell_{(4+n)}}\right)^{3+n}
\ .
\ee
\end{description}
We then conclude that GUP's obtained by linearly combining the quantum
mechanical expression with gravitational bounds do not imply
the holographic counting of degrees of freedom.
\section{Ng and Van~Dam GUP's}
\setcounter{equation}{0}
\label{ng}
An interesting GUP that satisfies the holographic principle in
four dimensions has been proposed by Ng and van~Dam in various
papers \cite{ngvd}. They start from the Wigner inequalities about
distance measurements with clocks and light signals.
\subsection{GUP in four dimensions}
\label{ng_4}
Suppose we wish to measure a distance $l$.
Our measuring device is composed of a clock, a photon detector and
a photon gun.
A mirror is put at the distance $l$ we want to measure and $m$ is
the mass of the system ``clock + photon detector + photon gun''.
We call ``detector'' the whole system and let $a$ be its size.
Obviously, we suppose
\be
a > r_{\rm g}\equiv\frac{2\,G_{\rm N}\, m}{c^2}=R_{\rm S}(m)
\ ,
\ee
which means that we are not using a black hole as a clock.
Be $\Delta x_1$ the uncertainty in the position of the detector.
Then the uncertainty on the detector velocity is
\be
\Delta v = \frac{\hbar}{2\,m\,\Delta x_1}
\ .
\ee
After a time $T = 2\,l/c$, elapsed during the light trip
detector--mirror--detector, the uncertainty on the detector position
(i.e.~the uncertainty on the actual length of the segment $l$) has
become
\be
\Delta x_{\rm tot} = \Delta x_1 + T\,\Delta v
= \Delta x_1 + \frac{\hbar\,T}{2\,m\,\Delta x_1}
\ .
\ee
We can minimize $\Delta x_{\rm tot}$ by suitably choosing
$\Delta x_1$,
\be
\frac{\partial \Delta x_{\rm tot}}{\partial \Delta x_1}
=0
\quad
\Rightarrow
\quad
(\Delta x_1)_{\rm min}=\left(\frac{\hbar\,T}{2\,m}\right)^{1/2}
\ .
\ee
Hence
\be
(\Delta x_{\rm tot})_{\rm min} =
(\Delta x_1)_{\rm min} + \frac{\hbar T}{2\,m\,(\Delta x_1)_{\rm min}}
=2\,\left(\frac{\hbar\,T}{2\,m}\right)^{1/2}
\ .
\ee
Since $T = 2\,l/c$, we have
\be
(\Delta x_{\rm tot})_{\rm min} =
2\,\left(\frac{\hbar\,l}{m\,c}\right)^{1/2}
\equiv\delta l_{\rm QM}
\ .
\label{dl_QM}
\ee
This is a purely quantum mechanical result obtained, for the first
time, by Wigner in 1958 \cite{Wig}.
From Eq.~(\ref{dl_QM}), it seems that we can reduce the error
$(\Delta x_{\rm tot})_{\rm min}$ as much as we want by choosing $m$
very large, since $(\Delta x_{\rm tot})_{\rm min}\to 0$ for
$m\to\infty$.
But, obviously, here gravity enters the game.
\par
A first remark is that the length $l$ must be greater than the
Schwarzschild radius of the detector with mass $m$,
\be
l>r_{\rm g} \quad \Rightarrow \quad
\frac{1}{m}>\frac{2\,G_{\rm N}}{l\,c^2}
\quad
\Rightarrow
\quad
(\Delta x_{\rm tot})_{\rm min}^2\gtrsim 8\,\ell_{\rm p}^2
\ .
\ee
A second consideration (due to Amelino-Camelia \cite{AC}) is that
the measuring device must not be a black hole,
\be
a>r_{\rm g} \quad \Rightarrow \quad
(\Delta x_{\rm tot})_{\rm min}^2 \gtrsim
8\,\ell_{\rm p}^2\,\frac{l}{a}
\ .
\ee
The typical scenario is $\ell_{\rm p} \ll a \leq l$.
Also in the ideal case, when $a \sim \ell_{\rm p}$, we have
\be
(\Delta x_{\rm tot})_{\rm min} \gtrsim 2\,\sqrt{2\,l\,\ell_{\rm p}}
\ .
\ee
\par
Ng and van~Dam have also considered a further source of error, a
purely gravitational error, besides the purely quantum mechanical
one already addressed.
Suppose the clock has spherical symmetry, with $a>r_{\rm g}$.
Then the error due to curvature can be computed from the
Schwarzschild metric surrounding the clock.
The optical path from $r_0>r_{\rm g}$ to a generic point $r>r_0$ is given
by (see, for example, Ref.~\cite{LL})
\be
c\,\Delta t=\int_{r_0}^{r}\frac{d \rho}{1-\frac{r_{\rm g}}{\rho}}
=(r-r_0) + r_{\rm g}\,\log\frac{r-r_{\rm g}}{r_0-r_{\rm g}}
\ ,
\ee
and differs from the ``true'' (spatial) length $(r-r_0)$.
If we put $a=r_0$, $l=r$, the gravitational error on the measure
of $(l-a)$ is thus
\be
\delta l_{\rm C} =
r_{\rm g}\,\log\frac{l-r_{\rm g}}{a-r_{\rm g}}
\sim r_{\rm g}\,\log\frac{l}{a}
\ ,
\ee
where the last estimate holds for $l > a \gg r_{\rm g}$.
\par
If we measure a distance $l\geq 2a$, then the error due to curvature
is
\be
\delta l_{\rm C} \geq r_{\rm g} \log 2 \simeq \frac{G_{\rm N} m }{c^2}.
\ee
Thus, according to Ng and van~Dam the total error is
\be
\delta l_{\rm tot} = \delta l_{\rm QM} + \delta l_{\rm C} =
2\,\left(\frac{\hbar\,l}{m\,c}\right)^{1/2} + \frac{G_{\rm N}\,m}{c^2}
\ .
\label{dl_tot4}
\ee
This error can be minimized again by choosing a suitable value for
the mass of the clock,
\be
\frac{\partial l_{\rm tot}(m)}{\partial m}=0
\quad
\Rightarrow
\quad
m_{\rm min}= c\,\left(\frac{\hbar\,l}{G_{\rm N}^2}\right)^{1/2}
\ee
and we then have
\be
\left(\delta l_{\rm tot}\right)_{\rm min} =
2\,\left(\frac{\hbar\,G_{\rm N}\,l}{c^3}\right)^{1/3} +
\left(\frac{\hbar\,G_{\rm N}\,l}{c^3}\right)^{1/3}
=3\,\left(\ell_{\rm p}^2\,l\right)^{1/3}
\ .
\label{otto}
\ee
The global uncertainty on $l$ contains therefore a term proportional
to $l^{1/3}$.
\subsection{GUP with $n$ extra dimensions}
\label{ng4+}
Ng and van~Dam's derivation can be generalized to the case with
$n$ extra dimensions.
The Wigner relation (\ref{dl_QM}) for the quantum mechanical error
is not  modified by the presence of extra dimensions and we just
need to estimate the error $\delta l_{\rm C}$ due to curvature.
\par
We are not considering now micro black holes created by the
fluctuations $\Delta E$ in energy, as in the previous Section.
Instead, we have to deal with (more or less) macroscopic clocks
and distances and this implies that we have to distinguish four
different cases:
\begin{enumerate}
\item
$0<L<r_{g}<a<l$;
\label{i}
\item
$0<r_{(4+n)}<L<a<l$;
\label{ii}
\item
$0<r_{(4+n)}<a<L<l$;
\label{iii}
\item
$0<r_{(4+n)}<a<l<L$;
\label{iv}
\end{enumerate}
where $r_{(4+n)}=R_{(4+n)}(m)$, and $r_{\rm g}=r_{(4)}$ as before.
The curvature error will be calculated (as in the previous
Subsection) by computing the optical path from $a\equiv r_o$ to
$l\equiv r$.
Of course, we will use a metric which depends on the relative
size of $L$ with respect to $a$ and $l$, that is
the usual four-dimensional Schwarzschild metric in the
region $L<r$, and the $4+n$ dimensional Schwarzschild
solution in the region $r<L$ (where the extra dimensions
play an actual role).
\par
In cases \ref{i}.~and \ref{ii}.~the optical path from $a$ to $l$
can be computed using just the four-dimensional Schwarzschild
solution and the result is given by Eq.~(\ref{otto}) in the
previous Subsection.
\par
In cases \ref{iii}.~and \ref{iv}.~we instead have to use the
Schwarzschild solution in $4+n$ dimensions \cite{MP},
\be
ds^2=-\left(1-\frac{C}{r^{n+1}}\right)c^2 dt^2
+ \left(1-\frac{C}{r^{n+1}}\right)^{-1}dr^2
+ r^2 d \Omega_{n+2}^2
\ ,
\ee
at least for a part of the optical path.
In the above,
\be
C=\frac{16\,\pi\,G_{(4+n)}\,m}{(n+2)\,A_{n+2}\,c^2}
\ ,
\ee
and $A_{n+2}$ is the area of the unit $(n+2)$-sphere, that is
\be
A_{n+2}=\frac{2\,\pi^{\frac{n+3}{2}}}{\Gamma \left(\frac{n+3}{2}\right)}
\ .
\ee
Besides, we note that, for $n=0$,
\be
C=\frac{2\,G_{\rm N}\,m}{c^2}=r_{\rm g}
\ ,
\ee
that is, $C$ coincides in four dimensions with the Schwarzschild
radius of the detector.
The Schwarzschild horizon is located where
$(1-C/r^{n+1})=0$, that is at $r=C^{1/(n+1)}\equiv r_{(4+n)}$, or
\be
r_{(4+n)}=
\left[
\frac{16\,\pi\,G_{(4+n)}\,m}{(n+2)\,A_{n+2}\,c^2}\right]^\frac{1}{n+1}
\ ,
\ee
in qualitative agreement with the expression obtained in
Subsection~\ref{gup_n}.
\par
In case \ref{iii}.~we obtain the optical path from $a$ to $l$ by
adding up the optical path from $a$ to $L$ and that from $L$ to $l$.
We have to use the solution in $4+n$ dimensions for the first part,
and the four-dimensional solution for the second part of the path,
\be
c\,\Delta t
&=&\int_a^L \left(1+\frac{C}{r^{n+1}-C}\right)\,dr
+\int_L^l \left(1+\frac{r_{\rm g}}{r-r_{\rm g}}\right)\,dr
\nonumber
\\
&=&(L-a)+(l-L)+ C\,\int_a^L \frac{dr}{r^{n+1}-C} + r_{\rm
g}\,\int_L ^l \frac{dr}{r-r_{\rm g}} \ . \ee We have shown before
that from $r_{(4+n)}<L$ (that holds in cases \ref{iii}.~and
\ref{iv}.) we can infer \be r_{g}<r_{(4+n)}<L \ . \ee Now, suppose
$a^{n+1} \gg C = r_{(4+n)}^{n+1}$, that is $a \gg r_{(4+n)}$, so
that we are not doing measures inside a black hole. Then
$r_{g}<r_{(4+n)}\ll a<L<l$ and
\be
c\,\Delta t& \simeq & (l-a)+
C\,\int_a ^L \frac{dr}{r^{n+1}} + r_{\rm g}\,\int_L ^l
\frac{dr}{r}
=(l-a)+\frac{C}{n}\,\left(\frac{1}{a^n}-\frac{1}{L^n}\right)
+r_{\rm g}\,\log\frac{l}{L} \nonumber
\\
&=&(l-a)+\frac{1}{n}\,\left(\frac{1}{a^n}-\frac{1}{L^n}\right)\,
\frac{16\,\pi\,G_{(4+n)}}{(n+2)A_{n+2}\, c^2}\,m
+\left(\frac{2\,G_{\rm N}}{c^2}\,\log\frac{l}{L}\right)\,m
\ .
\ee
The error caused by the curvature (when $a<L<l$) is therefore
linear in $m$,
\be
\delta l_{\rm C} =
\left[\frac{1}{n}\left(\frac{1}{a^n}-\frac{1}{L^n}\right)
\frac{16\,\pi\,G_{\rm N} L^n}{(n+2)\,A_{n+2}\,c^2} +
\frac{2\,G_{\rm N}}{c^2}\log\frac{l}{L}\right]\,m
\equiv K\,m
\ .
\ee
\par
We recall that the curvature error in four dimensions does not
contain the size of the clock.
On the contrary, this error in $4+n$ dimensions depends explicitly
on the size $a$ of the clock and on the size $L$ of the extra
dimensions.
Hence the total error is given by
\be
\delta l_{\rm tot} = \delta l_{\rm QM} + \delta l_{C} =
2\,\left(\frac{\hbar\,l}{m\,c}\right)^{1/2}+K\,m
=J\,m^{-1/2} + K\,m
\ ,
\ee
where $J=2\,(\hbar\,l/c)^{1/2}$ and $K$ was defined before.
This error can be minimized with respect to $m$,
\be
\frac{\partial \delta l_{\rm tot}}{\partial m}
=0
\quad \Rightarrow \quad
m_{\rm min}= \left(\frac{J}{2\,K}\right)^{2/3}
\ .
\ee
Finally,
\be
\left(\delta l_{\rm tot}\right)_{\rm min} &=&
\left(2^{1/3}+2^{-2/3}\right)\left(K\,J^2\right)^{1/3}
\nonumber
\\
&=&2\,\left(2^{1/3}+2^{-2/3}\right)\,
\left[\frac{1}{n}\,\left(\frac{1}{a^n}-\frac{1}{L^n}\right)\,
\frac{8\,\pi}{(n+2)\,A_{n+2}}\,\ell_{(4+n)}^{2+n}\,l
+\ell_{\rm p}^2\,l\,\log\frac{l}{L}\right]^{1/3}
\ ,
\label{sette}
\ee
where we used the definition of $J$ and $K$.
\par
In case \ref{iv}., the optical path from $a$ to $l$ can be
obtained by using simply the Schwarzschild solution in $4+n$
dimensions. We get \be c\,\Delta t =\int_a ^l
\left(1+\frac{C}{r^{n+1}-C}\right)\,dr = (l-a)+ C\,\int_a ^l
\frac{dr}{r^{n+1}-C} \ . \ee Suppose now, as before, that
$a^{n+1}\gg C=r_{(4+n)}^{n+1}$, that is $a \gg r_{(4+n)}$ (i.e.
our clock is not a black hole). We then have \be c\,\Delta t
\simeq (l-a)+ C\,\int_a ^l \frac{dr}{r^{n+1}} =(l-a) +
\frac{C}{n}\,\left(\frac{1}{a^n}-\frac{1}{l^n}\right) \ . \ee If
the distance we are measuring is, at least, of the size of the
clock ($l\geq 2\,a$), we can write \be c\,\Delta t \gtrsim (l-a) +
\frac{C}{n}\,\left(\frac{2^n-1}{2^n\,a^n}\right) \ . \ee The error
caused by the curvature is therefore (when $a<l<L$) \be \delta
l_{\rm C} = \frac{C}{n}\,\left(\frac{2^n-1}{2^n\,a^n}\right) \ .
\ee Here we again note that the curvature error in $4+n$
dimensions explicitly contains the size of the clock. The global
error can be computed as before \be \delta l_{\rm tot} = \delta
l_{\rm QM} + \delta l_{C} =2\,\left(\frac{\hbar
l}{m\,c}\right)^{1/2}
+\frac{C}{n}\,\left(\frac{2^n-1}{2^n\,a^n}\right) \ , \ee where
$C$ is linear in $m$. Minimizing $\delta l_{\rm tot}$ with respect
to $m$ can be done in perfect analogy with the previous
calculation. The result is \be \left(\delta l_{\rm
tot}\right)_{\rm min} =\left(2^{1/3}+2^{-2/3}\right)\,
\left(\frac{2^n-1}{2^n\,n}\,\frac{64\,\pi}{(n+2)\,A_{n+2}}\right)^{1/3}
\left(\frac{\ell_{(4+n)}^{n+2}\,l}{a^n}\right)^{1/3} \ .
\label{nove} \ee
\par
We note that the expression (\ref{sette}) coincides in the limit
$L\to a$ with Eq.~(\ref{otto}) (taking $l\geq 2\,a$), while, in
the limit $L\to l$, we recover from Eq.~(\ref{sette}) the expression
(\ref{nove}) (of course, supposing also that $l \geq 2\,a$).
\subsection{Holographic properties}
We finally examine the holographic properties of
Eq.~(\ref{nove}) for the GUP of Ng and van~Dam type in
$4+n$ dimensions.
We just consider the expression in Eq.~(\ref{nove}) because it
also represents the limit of Eq.~(\ref{sette}) for $L\to l$ and
$l\geq 2\,a$.
Moreover, for $n=0$, Eq.~(\ref{nove}) yields the four-dimensional
error given in Eq.~(\ref{otto}).
\par
Since we are just interested in the dependence of $n(V)$ on $l$ and
the basic constants, we can write
\be
\left(\delta l_{\rm tot}\right)_{\rm min}
\sim
\left(\frac{\ell_{(4+n)}^{n+2}\,l}{a^n}\right)^{1/3}
\ .
\ee
We then have that the number of degrees of freedom in the volume
of size $l$ is
\be
n(V)=\left(
\frac{l}{\left(\delta l_{\rm tot}\right)_{\rm min}}\right)^{3+n}
=\left(\frac{l^2\,a^n}{\ell_{\rm p}^2\,L^n}\right)^{1+\frac{n}{3}}
\ ,
\ee
and the holographic counting holds in four-dimensions ($n=0$)
but is lost when $n>0$.
Even if we take the ideal case $a \sim \ell_{(4+n)}$ we get
\be
n(V)=
\left(\frac{l}{\ell_{(4+n)}}\right)^{2\,\left(1+\frac{n}{3}\right)}
\ ,
\ee
and the holographic principle does not hold for $n>0$.
\section{Concluding remarks}
\setcounter{equation}{0}
\label{conc}
In the previous Sections, we have shown that the holographic
principle seems to be satisfied only by uncertainty relations
in the version of Ng and van~Dam and for $n=0$.
That is, only in four dimensions we are able to formulate
uncertainty principles which predict the same number of degrees
of freedom per spatial volume as the holographic counting.
This could be an evidence for questioning the existence of extra
dimensions.
Moreover, such an argument based on the holography could also be used
to support the compactification of string theory down to four
dimensions, given that there seems to be no firm argument which
forces the low energy limit of string theory to be four-dimensional
(except from the obvious observation of our world).
In this respect, we should also say that the cases \ref{iii}.~and
\ref{iv}.~of Subsection~\ref{ng4+} do not seem to have a good
probability to be realized in nature since, if there are extra
spatial dimensions, their size must be shorter than
$10^{-1}\,$mm \cite{Gund}.
Therefore, cases \ref{i}.~and \ref{ii}.~of Subsection~\ref{ng4+}
are more likely to survive the test of future experiments.
\par
A number of general remarks are however in order.
First of all, we cannot claim that our list of possible GUP's
is complete and other relations might be derived in different
contexts which accommodate for both the holography and extra
dimensions.
Further, one might find hard to accept that quantum
mechanics and general relativity enter the construction
of GUP's on the same footing, since the former is supposed to
be a fundamental framework for all theories while the latter
can be just regarded as a theory of the gravitational
interaction.
We might agree on the point of view that GUP's must be
considered as ``effective'' (phenomenological) bounds valid
at low energy (below the Planck scale) rather than ``fundamental''
relations.
This would in fact reconcile our result that four dimensions are
preferred with the fact that string theory (as a consistent
theory of quantum gravity) requires more dimensions through the
compactification which must occur at low energy, as we mentioned
above.
Let us also note that general relativity (contrary to usual
field theories) determines the space-time including the
causality structure, and the latter is an essential ingredient
in all actual measurements.
It is therefore (at least) equally hard to conceive uncertainty
relations which neglect general relativity at all.
This conclusion would become even stronger in the presence of
extra dimensions, since the fundamental energy scale of
gravity is then lowered \cite{add,RS} (possibly) within the scope of
present or near-future experiments and the gravitational radius
of matter sources is correspondingly enlarged \cite{Arg}.
\par
A final remark regards cases with less than four dimensions.
Since Einstein gravity does not propagate in such space-times
and no direct analogue of the Schwarzschild solution exists, one
expects a qualitative difference with respect to the cases
that we have considered here.
For instance, a point-like source in three dimensions would
generate a flat space-time with a conical singularity and no
horizon~\footnote{In three dimensions with negative cosmological
constant one also has the BTZ black hole which forms when two
point-like particles collide provided certain initial conditions
are satisfied.
For a recent review, see Ref.~\cite{Bir}.}.
Consequently, one does expect that the usual Heisenberg uncertainty
relations hold with no corrections for gravity.
\end{document}